\documentclass[a4paper,12pt]{article}
\pdfoutput=0
\usepackage{mathrsfs, amsfonts, amssymb, subeqn, graphicx, axodraw}
\usepackage{color,cite,hyperref,cancel,multirow}
\usepackage{bm,bbm}
\usepackage{bigdelim}

\long\def\rpl#1!!!#2!!!{\color[rgb]{0.7,0.2,0.2} $\blacktriangleright$
#1  
  \color[rgb]{0.2,0,1} #2 $\blacktriangleleft$ \color{black}}

\makeatletter

\@addtoreset{equation}{section}
\makeatother

\def\Eqn#1{Eq.\ (\ref{#1})}
\def\Eqs#1#2{Eqs.\ (\ref{#1}) and (\ref{#2})}
\def\3Eqs#1#2#3{Eqs.\ (\ref{#1}), (\ref{#2}) and (\ref{#3})}
\def\fig#1{Fig.~\ref{#1}}
\def\tabl#1{Table~\ref{#1}}
\def\sec#1{\S\,\ref{#1}}
\let\CapTion=\caption
\def\caption#1{\CapTion{\small\sf #1}}

\def\bg#1{\mathchoice
{{\mbox{\boldmath $#1$}}}
{{\mbox{\boldmath $#1$}}}
{{\mbox{\boldmath $\scriptstyle #1$}}}
{\mbox{\boldmath $\scriptscriptstyle #1$}}} 
\def\to{\longrightarrow}
\def\tbg#1{\tilde{\bg #1}}
\def\Re#1{\,{\rm Re}(#1)}
\def\Im#1{\,{\rm Im}(#1)}
\def\within#1to#2/{#1\mbox{ to }#2}
\def\lbullet#1,#2,#3,#4/{\Text(#1,#2)[c]{$\bullet$}\Line(#1,#2)(#3,#4)}
\def\textcite#1{Ref.~\cite{#1}}
\def\gap #1 #2/{\ensuremath{\it [#1,#2]}}

\newcommand{\tr}{\mathop{\rm Tr}}
\newcommand{\mod}{\mathrel{\rm mod}}
\def\e2spl{}
\def\diag{\mathop{\rm diag}}



\title{\bf Coxeter groups and the PMNS matrix}

\author{\bf Pritibhajan Byakti\\ 
\normalsize Department of Theoretical Physics \\ 
\normalsize Indian Association for the Cultivation of Science \\ 
\normalsize Kolkata 700 032, India
\\
\normalsize 
\and  
\bf  Palash B. Pal\\ 
\normalsize Saha Institute of Nuclear Physics\\ 
\normalsize 1/AF Bidhan-Nagar, Calcutta 700064, INDIA}

\date{}
\begin{document}

\maketitle

\begin{abstract}

We discuss symmetries of the Lagrangian of the leptonic sector.  We
consider the case when this symmetry group is a Coxeter group, and
identify the low energy residual symmetries with the involution
generators, i.e., generators with order equal to 2.  The number of
elements of the PMNS matrix predicted by this group structure would
depend on the number of generators of this group.  We analyze all
finite Coxeter groups with 2 to 4 generators and check which ones can
produce a PMNS matrix that is consistent with experimental data.  We
then extend the analysis to other groups which can be presented by
generators of order 2, and therefore can be seen as subgroups of
infinite Coxeter groups.

\end{abstract}

\section{Introduction}
Thanks to various experiments involving neutrino oscillations
\cite{Ardellier:2006mn, Ahn:2012nd, An:2013zwz,
  PhysRevLett.100.221803, Aartsen:2013jza}, the mixing matrix in the
leptonic sector is known to a good accuracy \cite{Capozzi:2013csa,
  Olive:2016xmw}.  The question then arises as to whether one can
understand the mixing matrix from some theoretical considerations
\cite{King:2013eh, Grimus:2011fk, Altarelli:2010gt}.  The
considerations would definitely involve some symmetry.  One can look
at the experimentally allowed values for the elements of the leptonic
mixing matrix, also called the PMNS matrix, and try to guess a
symmetry that would fit the values.  This approach has resulted, based
on experimental data available at different points of time, in various
schemes of neutrino mixing such as the bimaximal (BM)
\cite{Vissani:1997pa, Barger:1998ta} or the tribimaximal (TBM)
\cite{Harrison:2002er, Harrison:2002kp}.  To explain these patterns, a
different approach was considered, in which one writes down the
Lagrangian at some high energy scale like the grand unified scale, and
then predicts the patterns by considering the running of various
parameters down to low scales.  Literature of this approach is vast;
see Refs.~\cite{King:2013eh, Grimus:2011fk, Altarelli:2010gt} and
references therein.  There is yet another approach \cite{Lam:2007qc,
  Lam:2008sh, Lam:2009hn, Hernandez:2012ra, Hernandez:2012sk} in which
one starts with identifying symmetries of the mass terms in the
physical basis, i.e., the basis in which the mass matrices are
diagonal.  One assumes that these symmetries are remnants of a bigger
symmetry group that is presumably valid at some large energy scale.
One tries to identify this bigger group by starting with the remnant
symmetries as generators and building the group by imposing suitable
conditions on the generators.  The bigger symmetry then dictates the
structure of the PMNS matrix: if not the whole matrix, at least some
elements of it.  Luckily in this approach, one is not required to
write down any Lagrangian for high energy scale.  Simple
group-theoretical considerations give us the PMNS matrix.

We assume that the neutrinos are Majorana particles, in which case the
symmetries associated with the neutrino field redefinitions must be
involutions, i.e., the square of the symmetry transformations must
yield the identity transformation.  In the charged lepton sector, the
redefinitions can be different.  For the sake of simplicity, one
assumes a discrete symmetry in this sector as well \cite{Lam:2007qc}.
At first, people tried to obtain the TBM matrix with this assumption,
because the TBM was consistent with experimental data at that time.
When the experimental data confirming a non-zero value of
$\theta_{13}$ poured in \cite{Ahn:2012nd, An:2013zwz}, the work
continued with the same philosophy, but with the changed data.  In
such attempts \cite {Hernandez:2012ra, Hernandez:2012sk}, it was first
assumed that the eigenvalues of the symmetry generators of the charged
leptons are non-degenerate.  However, degeneracy cannot be ruled out
from any physical   consideration, and a large number of
authors have explored various symmetries, with
or without degenerate eigenvalues \cite{deAdelhartToorop:2011re,
  Holthausen:2012wt, Hagedorn:2013nra, Lavoura:2014kwa,
  Fonseca:2014koa, Hu:2014kca, Ge:2014mpa, Ding:2014ora,
  Turner:2015uta}.  They have identified several groups that can
predict at least some of the elements of the PMNS matrix.

In this work, we perform a systematic analysis of the possibility that
all symmetries in the leptonic sector of the Lagrangian are generated
by involutions, i.e., elements which are of order 2.  A class of these
groups are called Coxeter groups, a term that will be defined in
\sec{s:fc}, where such groups will be discussed in detail.  Some
groups of this kind were arrived at by Lam \cite{Lam:2014kga} while
searching for ``built-in'' symmetries in SO(10) grand unified models.
In addition, many of the groups that we encounter in our search have
been discussed sporadically by other authors (see Table II of
Ref.~\cite{Altarelli:2010gt} and references therein), and therefore
many of the results that we obtain are not new.  The novelty of our
approach is the use of the theory of Coxeter groups.  We use the
theory to identify all Coxeter groups with 4 or fewer generators, and
discuss their relevance in this context.  Further, it also helps us
identifying other involution groups, because they must be subgroups of
infinite Coxeter groups.

When one deals with groups generated by involutions only, degeneracy
in the eigenvalues cannot be avoided since there exist only two
distinct eigenvalues for each generator.  Therefore, after a general
analysis of the symmetries of the mass terms and currents and the
presentation of some earlier results in \sec{s:sym}, we discuss how to
tackle degenerate phases in \sec{s:deg}.  We summarize the
experimental data in \sec{s:exp} and set up the strategy for the
search of groups that might be consistent with the data.  In
\sec{s:fc} we introduce Coxeter groups and perform the analysis with
finite Coxeter groups with 4 or less generators.   In \sec{s:inv} we
consider other finite groups generated by involutions only, by
treating them as subgroups of infinite Coxeter groups, as argued
before.  We summarize our findings and our outlook in \sec{s:out}.

\section{Consideration of symmetries}\label{s:sym}
We start with a brief review of the method \cite{Hernandez:2012ra,
  Hernandez:2012sk}.  If the Lagrangian of the leptonic sector is
written in terms of the mass eigenstates, it contains the following
terms:
\begin{eqnarray}
\mathscr L = \sum_{\ell,\alpha} \left[ \frac{g}{\surd2} \left(
  \bar\ell U_{\ell\alpha} \gamma^\mu L \nu_\alpha W_\mu^+ + \mbox{h.c.}
  \right) - M_\ell \bar \ell \ell - \frac12 m_\alpha \nu_\alpha^\top C
\nu_\alpha \right] \,,
\label{lag}
\end{eqnarray}
where $\ell=e,\mu,\tau$, whereas the neutrino eigenstates are indexed
by $\alpha=1,2,3$.  Note that here and anywhere else, we use the
Einstein summation convention for Lorentz indices, but not for indices
used for differentiating different particles.  The neutrinos have been
assumed to be Majorana particles.  There are of course other terms in
the Lagrangian that involves the leptons, but they are not crucial for
the argument that follows.

The mass terms in \Eqn{lag} admit the following symmetries:
\begin{subequations}
\label{masssym}
\begin{eqnarray}
\nu_\alpha &\to& \eta_\alpha \nu_\alpha \,, \\*
\ell &\to& e^{i\phi_\ell} \ell \,.
\end{eqnarray}
\end{subequations}
Because each neutrino eigenstate has to satisfy the Majorana
condition,
\begin{eqnarray}
\bar\nu_\alpha = \nu^\top C
\end{eqnarray}
for some unitary matrix $C$, each $\eta_\alpha$ must be real, i.e.,  
\begin{eqnarray}
\eta_\alpha = \pm 1 \qquad \forall \alpha \,.
\end{eqnarray}
The maximum possible symmetry in \Eqn{masssym} is therefore $[Z_2]^3$
from the neutrino sector and $\rm [U(1)]^3$ from the charged leptons.
The interaction of leptons with the $Z$-boson and the Higgs boson also
obey this symmetry, which is why we have not written them down in
\Eqn{lag}.  On the other hand, the charged current interaction term of
\Eqn{lag} remains invariant if we augment the transformations of
\Eqn{masssym} by the rule
\begin{eqnarray}
U_{\ell\alpha} \to U_{\ell\alpha} e^{-i\phi_\ell} \eta_\alpha \,.
\label{Usymm}
\end{eqnarray}
This is then the arbitrariness in the definition of the PMNS matrix
$U$.

Because of this arbitrariness, it would be pointless to try to predict
the PMNS matrix $U$.  However, one can try whether one can somehow use
the symmetry of \Eqn{Usymm} to get some information about the absolute
values of the matrix elements of $U$.  To this end, we rewrite
\Eqn{masssym} in the form
\begin{subequations}
\label{matsym}
\begin{eqnarray}
\bg\nu &\to& S \bg\nu \,, \\*
\bg\ell &\to& T \bg\ell \,,
\end{eqnarray}
\end{subequations}
where the bold letters indicate column matrices in flavor space, whereas
$S$ and $T$ are diagonal matrices.  Now, in order to make the problem
more tractable, one assumes that
\begin{eqnarray}
\det S = \det T = 1 \,,
\label{det=1}
\end{eqnarray}
so that
\begin{subequations}
\label{phasereln}
\begin{eqnarray}
\eta_1 \eta_2 \eta_3 &=& 1 \,, 
\label{etareln} \\* 
\phi_e + \phi_\mu + \phi_\tau &=& 0 \mod 2\pi.
\label{phireln}
\end{eqnarray}
\end{subequations}
Then the symmetry for the neutrino fields can be generated by the
matrices 
\begin{eqnarray}
S_1 = \diag (1,-1,-1), \qquad 
S_2 = \diag (-1,1,-1), \qquad 
S_3 = \diag (-1,-1,1).
\label{S}
\end{eqnarray}
Since 
\begin{eqnarray}
S_1 S_2 S_3 = 1 \,, 
\label{Sprod}
\end{eqnarray}
there are only two independent operations, and therefore the symmetry
is at most $Z_2 \times Z_2$.  On the other hand, the symmetry in the
charged lepton sector, already reduced to $\rm [U(1)]^2$ because of
\Eqn{phireln}, is further reduced by assuming that $T$ is the
representation of a discrete group.  The discrete group must then be
of the form $Z_n \times Z_{n'}$.

Let us now discuss the whole thing from the flavor basis of
neutrinos.  The neutrino fields in this basis are given by
\begin{eqnarray}
\tilde{\bg \nu} = U \bg \nu \,,
\end{eqnarray}
with the tilde indicating the basis in which the charged lepton mass
matrix is diagonal.  The neutrino mass matrix in this basis takes the
form
\begin{eqnarray}
\tilde{\bg m} = U^* \bg m U^\dagger \,.
\end{eqnarray}
The symmetry of \Eqn{matsym} now becomes
\begin{eqnarray}
\tilde{\bg\nu} \to S' \tilde{\bg\nu} \,, \qquad 
\tbg m \to S'^\top \tbg m S'
\label{S'symm}
\end{eqnarray}
where
\begin{eqnarray}
S' = U S U^\dagger \,,
\label{S'}
\end{eqnarray}
$S$ being any of the matrices of \Eqn{S}.  There is no reason to
assume that $S'$ and $T$ commute.  Depending on their structures, we
can deduce the moduluses of different elements of the PMNS matrix $U$.

As the simplest example, we can consider that the flavor groups in
neutrino and charged lepton sectors each has one generator only.
Since \Eqn{S'} implies that $S'^2=1$, it means that the symmetry group
will be $Z_2$ for the neutrino sector and $Z_n$, for some $n$, in the
charged lepton sector.  The $Z_2$ group of the neutrino sector must be
generated by one of the matrices $S'_\alpha$, defined through
\Eqn{S'}.  One can then introduce the group element
\begin{eqnarray}
W_\alpha = S'_\alpha T \,,
\label{W}
\end{eqnarray}
and assume that this element also has a finite order.  The group now
will be defined by the relations
\begin{eqnarray}
S_\alpha'^2 = T^n = W_\alpha^p = 1 \,,
\label{vonDyck}
\end{eqnarray}
which defines the von Dyck group $D(2,n,p)$.

In order to obtain the elements of the PMNS matrix, one evaluates 
\begin{eqnarray}
a_\alpha \equiv \tr (W_\alpha) = 
\sum_\ell (2|U_{\ell\alpha}|^2 -1)e^{i \phi_\ell}.
\label{trW}
\end{eqnarray}
Using real and imaginary parts of the above equation with unitary
condition for $U$, one can exactly solve the absolute values of one
column of the PMNS matrix.  The solution is given
by~\cite{Hernandez:2012ra, Hernandez:2012sk}
\begin{eqnarray}
|U_{e\alpha}|^2 & =&  \frac{ \Re{a_\alpha} \cos \frac{\phi_e}{2} + 
\cos\frac{3\phi_e}{2} - \Im{a_\alpha} \sin \frac{\phi_e}{2} }
{4\sin{\frac{\phi_e-\phi_\mu}{2}} \sin{\frac{\phi_\tau-\phi_e}{2}}}\,,
\label{Uealpha} 
\end{eqnarray}
with $|U_{\mu\alpha}|^2$ and $|U_{\tau\alpha}|^2$ obtained by making
cyclic permutation of the indices $e$, $\mu$, $\tau$.  Similarly, if
we have two $Z_2$ symmetries in the neutrino sector, we can obtain
expressions for another column.  Hence, by using unitarity conditions,
the remaining column can also be determined, which means that we would
know the absolute values of all entries of the PMNS matrix.

\section{Degenerate phases}\label{s:deg}
An expressions like that in \Eqn{Uealpha} cannot be the most general
formula for determination of the PMNS matrix elements
\cite{Lavoura:2014kwa}.  One can easily see that they break down when
any two of the phases $\phi_e, \phi_\mu$ and $\phi_\tau$ are
equal.  There can be three possibilities with a two-fold degeneracy:
\begin{eqnarray}
T_e &=& \diag \Big(e^{-2 i \phi_e},e^{i \phi_e},e^{ i \phi_e} \Big)
\,,\nonumber\\*
T_\mu &=& \diag \Big( e^{i \phi_\mu},e^{-2 i \phi_\mu},e^{ i
  \phi_\mu} \Big) \,,
\nonumber\\*
T_\tau &=& \diag \Big(e^{i \phi_\tau},e^{i \phi_\tau},e^{- 2 i
  \phi_\tau} \Big) \,.
\label{T}
\end{eqnarray}

In analogy with \Eqs{W}{trW}, we now define the following quantities:
\begin{subequations}
\label{Wa}
\begin{eqnarray}
W_{\ell\alpha} &=& T_\ell S'_\alpha \,, 
\label{Wlalpha}\\
a_{\ell\alpha} &=& \tr (W_{\ell\alpha}) = \tr (T_\ell U S_\alpha
U^\dagger) \,.
\label{alalpha}
\end{eqnarray}
\end{subequations}
The traces are easily determined and one obtains
\begin{eqnarray}
a_{\ell\alpha} = - e^{-2i \phi_\ell} + 2 |U_{\ell\alpha}|^2
(e^{-2i \phi_\ell} - e^{i\phi_\ell}) \,.
\label{aU}
\end{eqnarray}
For generators of order 2, $\phi_\ell=\pi$.  This fact has several
important consequences.  First, we see from \Eqn{T} that now the
generators $T_\ell$ are of the form
\begin{eqnarray}
T_e = \diag (1,-1,-1), \qquad 
T_\mu = \diag (-1,1,-1), \qquad 
T_\tau = \diag (-1,-1,1) \,,
\label{T2}
\end{eqnarray}
so that
\begin{eqnarray}
T_e T_\mu T_\tau = 1 \,,
\label{Tprod}
\end{eqnarray}
implying that only two of these generators are independent.  Second,
we can now use \Eqn{aU} to write
\begin{eqnarray}
|U_{\ell\alpha}|^2 = \frac14 \Big(1 + a_{\ell\alpha}\Big) \,,
\label{aUcox}
\end{eqnarray}
which shows that $a_{\ell\alpha}$ will have to be real.

Note that $W_{\ell\alpha}$, defined in \Eqn{Wlalpha}, must be an
element of the symmetry group.  Suppose that the order of this element
is $p_{\ell\alpha}$.  The quantity $a_{\ell\alpha}$, being the trace
of $W_{\ell\alpha}$, is therefore nothing but the sum of three
eigenvalues of $W_{\ell\alpha}$, i.e., of three $p_{\ell\alpha}^{\rm
  th}$ roots of unity.  Let us denote these three roots by
$e^{i\theta_1}$, $e^{i\theta_2}$ and $e^{i\theta_3}$, where each of
these $\theta$'s is of the form $2\pi k/p_{\ell\alpha}$, with possibly
different integral values of $k$ but the same value of
$p_{\ell\alpha}$ that is the order of the element $W_{\ell\alpha}$.
Since by \Eqn{det=1} the determinant is unity, we obtain
\begin{eqnarray}
\theta_1 + \theta_2 + \theta_3 = 0 \mod 2\pi \,.
\label{sumtheta}
\end{eqnarray}
Further, since the sum of the roots is real, we have
\begin{eqnarray}
\sin\theta_1 + \sin\theta_2 + \sin\theta_3 = 0 \,.
\label{sumsine}
\end{eqnarray}
Using \Eqn{sumtheta} to eliminate $\theta_3$, we can write
\Eqn{sumsine} as 
\begin{eqnarray}
\sin\theta_1 + \sin\theta_2 = \sin(\theta_1+\theta_2) \,,
\end{eqnarray}
which can be rewritten in the form
\begin{eqnarray}
\tan {\theta_1 \over 2} = - \tan {\theta_2 \over 2} \,.
\end{eqnarray}
The most general solution of this equation is
\begin{eqnarray}
\theta_1 = 2\pi m - \theta_2 
\end{eqnarray}
for some integer $m$.  Using this relation along with \Eqn{sumtheta},
we can determine all three eigenvalues of $W_{\ell\alpha}$ in terms of
one parameter, and write
\begin{eqnarray}
a_{\ell\alpha} &=& \exp \left( 2\pi i {k_{\ell\alpha} \over
  p_{\ell\alpha}} \right) + \exp \left( - 2\pi i {k_{\ell\alpha} \over
  p_{\ell\alpha}} \right) + 1 \nonumber \\ 
&=& 1 + 2 \cos \left( 2\pi {k_{\ell\alpha} \over
  p_{\ell\alpha}} \right) \,,
\label{rootsum}
\end{eqnarray}
restoring the definition of the $\theta$'s.  We can then use
\Eqs{aUcox}{rootsum} to obtain
\begin{eqnarray}
|U_{\ell\alpha}|^2 = \frac12 \left[ 1 + \cos  \Big({2\pi  
  \frac{k_{\ell\alpha}}{p_{\ell\alpha}}} \Big) \right]  =
\cos^2 \Big({\pi  
  \frac{k_{\ell\alpha}}{p_{\ell\alpha}}} \Big)  \,.
\label{Usoln}
\end{eqnarray}

The task is now to find different combinations of $p_{\ell\alpha}$ and
$k_{\ell\alpha}$ that will produce values of $|U_{\ell\alpha}|^2$ that
fall within the experimentally allowed ranges.  Before embarking on
this journey, we summarize the experimental results that we are trying
to fit.

\section{Confronting experimental results}\label{s:exp}
The PMNS matrix is written in terms of three angles and three
CP-violating phases~\cite{Chau:1984fp}:
\begin{eqnarray}
U &=& \left( \begin{array}{ccc}
c_{12} c_{13} & s_{12} c_{13} & s_{13} e^{-i\delta} \\ 
-s_{12} c_{23} - c_{12} s_{23} s_{13} e^{i\delta} & c_{12} c_{23} -
s_{12} s_{23} s_{13} e^{i\delta} & s_{23} c_{13} \\
s_{12} s_{23} - c_{12} c_{23} s_{13} e^{i\delta} & - c_{12} s_{23} -
s_{12} c_{23} s_{13} e^{i\delta} & c_{23} c_{13} \\
\end{array}\right) \nonumber \\*
&& \hspace*{0.4\textwidth} \times \diag (1, e^{i \beta_2},
e^{i\beta_3}) \,, 
\end{eqnarray}
where, for example, $c_{12} = \cos\theta_{12}$ and $s_{12} =
\sin\theta_{12}$.  There is no limit on the phases $\beta_2$,
$\beta_3$ and $\delta$ at the $3\sigma$ level.  The $3\sigma$ limits
on the other parameters are as follows~\cite{Capozzi:2013csa,
  Olive:2016xmw}:
\begin{eqnarray}
\begin{array}{l|cc|cc}
\multirow{3}{*}{\mbox{Parameter}}
& \multicolumn{4}{c}{3\sigma \mbox{ limits for}} \\ \cline{2-5}
& \multicolumn{2}{c|}{\mbox{Normal hierarchy}} 
& \multicolumn{2}{c}{\mbox{Inverted hierarchy}} \\ \cline{2-5}
& \mbox{Lower limit} & \mbox{Upper limit} 
& \mbox{Lower limit} & \mbox{Upper limit} \\ \hline 
\sin^2 \theta_{12} & 0.250 & 0.354 & 0.259 & 0.359 \\ 
\sin^2\theta_{23}  & 0.379 & 0.616 & 0.383 & 0.637 \\
\sin^2\theta_{13}  & 0.0185 & 0.0246 & 0.0186 & 0.0248 \\
\end{array}
\end{eqnarray}
Using these limits, we find $3\sigma$ limits on the absolute values of
the elements of the PMNS matrix.  For normal hierarchy (NH), the
limits are as follows,
\begin{subequations}
\label{Usqlim}
\begin{eqnarray}
|U^2| = \left( \begin{array}{ccc}
\within 0.630 to 0.736/ & \within 0.244 to 0.347/ & \within 0.0185 to
0.0246/ \\ 
\within 0.0432 to 0.299/ & \within 0.180 to 0.532/ & \within 0.370 to
0.604/ \\ 
\within 0.0403 to 0.295/ & \within 0.180 to 0.530/ & \within 0.375 to
0.609/ 
\end{array}\right) \,,
\label{NHlim}
\end{eqnarray}
whereas for the inverted hierarchy (IH), the limits are slightly
different: 
\begin{eqnarray}
|U^2| = \left( \begin{array}{ccc}
\within 0.630 to 0.736/ & \within 0.244 to 0.347/ & \within 0.0186 to
0.0248/ \\ 
\within 0.0389 to 0.298/ & \within 0.168 to 0.529/ & \within 0.374 to
0.625/ \\ 
\within 0.0409 to 0.302/ & \within 0.182 to 0.546/ & \within 0.354 to
0.605/ 
\end{array}\right) \,.
\label{IHlim}
\end{eqnarray}
\end{subequations}
It should be noted that the ranges indicated here do {\em not} pertain
to values of the matrix elements of $U^2$.  Rather, each entry denotes
the range of the range of modulus squared of an element of the PMNS
matrix $U$.  These are the ranges that we will use for checking the
feasibility of getting a particular Coxeter group.

In the approach that we are going to take, all elements of the PMNS
matrix cannot be found in general.  We will discuss, depending on a
particular choice of the group, how many elements of the PMNS matrix
can be predicted, and will check how they fare in the light of
experimental data.

In this pursuit, we will use \Eqn{Usoln}.  In order to avoid
double counting and unnecessary work, it is useful to keep the
following guidelines in mind.
\begin{enumerate}
\item The solution $k_{\ell\alpha}=0$ is not allowed for any
  $p_{\ell\alpha}$, because it gives $|U_{\ell\alpha}|^2=1$ which is
  not allowed for any element of the PMNS matrix.

\item If $p_{\ell\alpha}$ is even, the value $k_{\ell\alpha}=\frac12
  p_{\ell\alpha}$ is not allowed as well, because it gives
  $U_{\ell\alpha}=0$, which is unacceptable for any element of the
  PMNS matrix.

\item Values of $k_{\ell\alpha}$ with
\begin{eqnarray}
k_{\ell\alpha} > \frac12 p_{\ell\alpha}
\end{eqnarray}
do not produce any new value of $|U_{\ell\alpha}|^2$ that is not
already encountered with smaller values of $k_{\ell\alpha}$.  Hence,
these are irrelevant for our search.

\item Since only the ratio of $k_{\ell\alpha}$ and $p_{\ell\alpha}$
  appears in \Eqn{Usoln}, any common factor in the two numbers is
  irrelevant. 

\end{enumerate}
Combining these guidelines, we see that we only need to check for the
values 
\begin{eqnarray}
0 < k_{\ell\alpha} < \frac12 p_{\ell\alpha} \,,
\label{klimit}
\end{eqnarray}
with 
\begin{eqnarray}
\gcd (k_{\ell\alpha}, p_{\ell\alpha}) = 1 \,.
\label{gcd}
\end{eqnarray}
For $p_{\ell\alpha} \leq 5$, we present the result of these
checks.
\begin{eqnarray}
\begin{array}{cccl}
\hline
\multirow{2}{*}{$p_{\ell\alpha}$}  &
\multirow{2}{*}{$k_{\ell\alpha}$} &
\multirow{2}{*}{$|U_{\ell\alpha}|^2$} & 
\mbox{$\ell\alpha$ combinations that} \\
&&& \mbox{give $|U_{\ell\alpha}|^2$ in range}  \\ 
\hline
2  & 1 & 0 &  {\rm none}  \\
3  & 1 & \frac14 & e2, \mu1, \mu2, \tau1, \tau2 \\
4  & 1 & \frac12 & \mu2, \mu3, \tau2, \tau3  \\ 
5  & 1 & \frac18(3+\surd5) & e1 \\ 
5  & 2 & \frac18(3-\surd5) & \mu1, \tau1 \\ 
\hline
\end{array}
\label{allowed}
\end{eqnarray}
Fortunately, the availability of the solutions is not sensitive to the
small differences of the allowed values that appear in \Eqn{NHlim} and
in \Eqn{IHlim}, so our subsequent analysis apply equally well for both
hierarchies.

If the number of generators of the group is more than 2, there is a
different kind of relation that we will need to satisfy.  To
understand the point, let us assume that we have a group with two
$S'$-type and one $T$-type involution generators.  There will be two
different combinations $W_{\ell\alpha}$, and therefore two elements of
the same column will be determined through \Eqn{Usoln}.  However, it
is important to notice that, using \Eqs{Sprod}{S'}, and the unitarity
of the PMNS matrix $U$, we can write
\begin{eqnarray}
S'_1 S'_2 S'_3 = 1 \,, 
\label{S'prod}
\end{eqnarray}
which means that even the third $S'$-type matrix is also an element of
the group.  Instead of taking $S'_1$ and $S'_2$, say, as the
generators, we could have also chosen $S'_1$ and $S'_3$, along with
the $T$-type generator.  If we had done that, the modulus of the third
element of the column of the PMNS matrix would also have been
determined by a relation of the form given in \Eqn{Usoln}.  The
unitarity condition on the three elements of the same column would
have then ensured that
\begin{eqnarray}
\sum_{\alpha=1,2,3} \cos^2 \Big( \pi  
  {k_{\ell\alpha} \over p_{\ell\alpha}} \Big) = 1 \,.
\label{sumcosS}
\end{eqnarray}
Similarly, if we had considered a group with two $T$-type and one
$S'$-type involution generators, we would have obtained
\begin{eqnarray}
\sum_{\ell=e,\mu,\tau} \cos^2 \Big( \pi  
  {k_{\ell\alpha} \over p_{\ell\alpha}} \Big) = 1 \,.
\label{sumcosT}
\end{eqnarray}
Both kinds of equations are of the same form,
\begin{eqnarray}
\cos^2 \Big( \frac{\pi n_1}{N} \Big) + \cos^2 \Big(
  \frac{\pi n_2}{N} \Big) + \cos^2 \Big(\frac{\pi n_3}{N} \Big) = 1 \,,
\label{n/N}
\end{eqnarray}
with a suitably defined $N$ which can be the LCM of the numbers
$p_{\ell\alpha}$.  If we consider a group with four generators, both
\Eqs{sumcosS}{sumcosT} will apply, and therefore there will be six
equations of the form given in \Eqn{n/N} that need to be satisfied,
one for each row and for each column.

There are trivial solutions to \Eqn{n/N} in which at least one of the
cosines is zero.  For example, if $N$ is even, we have solutions in
which one of the $n_i$'s is equal to zero and the other two equal to
$N/2$, i.e., one of the cosine-squared values is equal to 1 and the
other zero.  Or we can have solutions like
\begin{eqnarray}
n_1 + n_2 = \frac12 N \,, \qquad n_3 = \frac12 N \,,
\label{trivial}
\end{eqnarray}
along with permutations of the set of numbers $n_1$, $n_2$ and $n_3$.
Such solutions will not be useful for us, because they would imply
zeroes as elements of the PMNS matrix, which are untenable by
experimental results.  We need to find solutions of \Eqn{n/N} where
each of the cosines is non-zero.  It can be analytically shown that
\cite{Fonseca:2014koa, Hu:2014kca}, subject to the guidelines
summarized in \Eqs{klimit}{gcd}, the only solutions of \Eqn{n/N} are
these:
\begin{eqnarray}
\begin{array}{cll}
\hline
N & \{n_1, n_2, n_3\} & \mbox{Values of $|U_{\ell\alpha}|^2$} \\
\hline
12 & \{3, 4, 4 \} & \{ \frac12, \frac14, \frac14 \} \\
15 & \{3, 5, 6 \} & \{ \frac18(3+\surd5), \frac14, \frac18(3-\surd5) \} \\
\hline
\end{array}
\label{2solns}
\end{eqnarray}
For each choice of $N$, we have also given, in the last column, the
mod-squared values of the entries of the PMNS matrix in the row or
column for which that value of $N$ applies.

One interesting point to note is that  one cannot obtain the TBM
form if the generators of the flavor group are involutions.  The
reason is that the \Eqn{Usoln} shows that the absolute-squared values
of each element must be of the form of the cosine-squared of an angle
which is a rational multiple of $\pi$, and the TBM form contains
absolute-squared values equal to $\frac13$, $\frac16$ etc which are
not.  Of course, one can obtain TBM form when not all $T$-type and
$S$-type generators are taken as involutions \cite{Altarelli:2010gt,
  Hagedorn:2006ug, Bazzocchi:2008ej, Bazzocchi:2009pv, Meloni:2009cz,
  Ding:2009iy, Morisi:2010rk, Hagedorn:2010th}.  This issue of
arbitrariness in the choice of generators will be elaborated in
\sec{s:fcd}.  The BM matrix can be obtained with involution
generators, but we will not consider it further since it contains a
zero element, just as the TBM does.

\section{Requirement of irreducible representations}\label{s:irrep}
Experimental data show that none of the elements of $U$ is zero.  This
fact has an important implication on the $T$ and $S'$ generators, as
we show now.
 
In the basis in which the $T$ generators are diagonal, the relation
between the $S'$ generators and the PMNS matrix can be read from
\Eqn{S'}:
\begin{eqnarray}
\Big( S'_\alpha \Big)_{\ell\ell'} 
= \cases{ 2 U_{\ell\alpha} U^*_{\ell'\alpha} - 1 & if $\ell=\ell'$, \cr
2 U_{\ell\alpha} U^*_{\ell'\alpha} & otherwise,
} 
\label{S'll'}
\end{eqnarray}
using the $S$ generators given in \Eqn{S}.  This shows that the
off-diagonal elements of $S'_\alpha$ cannot be zero, and therefore
$S'_\alpha$ cannot be block diagonal.  Conversely, if $S'_\alpha$ has
to be block diagonal, some of its off-diagonal elements must vanish,
requiring some elements of $U$ to vanish.  Thus, the statement that
all elements of $U$ are non-zero is equivalent to the statement that
the $S'$ generators are not block diagonal in the representation in
which the $T$ generators are.  Therefore, the representation
comprising the $T$ and $S'$ generators has to be an irreducible
representation (irrep) in order that all elements of the PMNS matrix
are non-zero.  Since we are dealing with three generations of
fermions, it means that the flavor group must have 3-dimensional
irreps.

The point can be made in another way \cite{thankref}.  Suppose we take
a set of matrices $M_I$ and try to find a matrix $M$ that commutes
with each matrix in the set.  Obviously, the unit matrix, or any
multiple of it, will be solutions to the problem.  If there is no
other solution, then by Schur's theorem the matrices are irreducible.
If we only have two different $T$'s of \Eqn{T2} in the set $M_I$, then
it is straight forward to show that the general solution for $M$ is a
diagonal matrix.  If now we also put one of the $S'$ matrices in the
collection $M_I$, then $M$ can only be a multiple of the identity
matrix provided $S'$ has no zero element.  Hence, with two $T$-type
and one $S'$-type generators, we need 3-dimensional irreps.
The same is true if we have to generators of the $S'$-type and one of
the $T$-type, and the proof is the same if we change over to the basis
in which the $S'$ generators are diagonal.

If, however, we have only one $T$-type and one $S'$-type generators,
that is not the case.  Without loss of generality, let us say that the
set $M_I$ contains the $T$-type generator $T_e$.  The most general $M$
that commutes with it has the form
\begin{eqnarray}
M = \left( \begin{array}{ccc} 
a & 0 & 0 \\ 
0 & m_{22} & m_{23} \\ 
0 & m_{32} & m_{33} 
\end{array} \right) \,.
\end{eqnarray}
  Suppose
now we put the requirement that this $M$ should commute with $S'_1$.
That would require
\begin{eqnarray}
m'_{12} = m'_{13} = m'_{21} = m'_{31} = 0 \,,
\end{eqnarray}
where these are elements of the matrix $M' = U^\dagger M U$.  The four
zeros will give four homogeneous equations for the quantities
$m_{22}-a$, $m_{23}$, $m_{32}$ and $m_{33}-a$.  Using the unitarity of
the PMNS matrix $U$, one can easily show that the determinant of the
co-efficient matrix vanishes, implying that non-zero solutions are
possible.  This means that the 3-dimensional matrices are reducible,
implying that a block-diagonal solution for $S'$ can be obtained.
That would give zero elements in $U$, as expected from \Eqn{S'll'}.
In fact, later when we discuss groups with two generators, we show
explicitly that those are groups which do not have any 3-dimensional
irreducible representation.

Thus, if we encounter a group that has only 1 and 2 dimensional
irreps, it is useless for us.  A group that admits a 3 dimensional
irrep is certainly fine.  If a group does not have 3 dimensional irrep
but has higher dimensional irreps, it also cannot be ruled out.  The
reason is the following.  We are going to follow the Coxeter diagrams,
which imply the number of generators and some relations between them,
as exemplified in \Eqn{prexamp}.  Thus, the groups that we will find
will be the ones which are consistent with some conditions on the
generators.  Suppose a given set of conditions on a fixed number of
generators specifies a group $G$ if we assume that there are no other
condition connecting the generators.  If we impose an extra condition,
we will still obtain a group $H$ which will be a subgroup of $G$.
Even if $G$ does not have a 3-dimensional irrep, it is possible that
$H$ does, and maybe this subgroup is responsible for the structure of
the PMNS matrix.  Thus, a flavor group can be acceptable provided it
has irreps of dimension 3 or more.

It is to be understood that this constraint has nothing to do with the
specific choice of the flavor group.  \Eqn{S'll'} holds irrespective
of the underlying group structure: it just says that, in a basis in
which the $T$-generators are diagonal, the matrix $U$ diagonalizes the
matrices $S'_\alpha$.  Thus, this constraint has to be obeyed in any
model of this form, irrespective of the $T$ and $S'$-type generators.
The condition that is crucial for this conclusion is the absence of
zeroes in the PMNS matrix, as we have mentioned at the beginning of
this section.

We want to emphasize that our discussion pertains to the case where the  
flavor symmetries are exact. If the symmetry is broken, explicitly or
spontaneously, then symmetry breaking terms can contribute to the PMNS
matrix as well and give a form that is not block-diagonal.  There are
discussions of such scenarios in the literature \cite{Blum:2007jz}.
We do not discuss this possibility.

\section{Finite Coxeter groups}\label{s:fc}
\subsection{Coxeter groups and Coxeter diagrams}\label{s:fcd}
In \sec{s:deg}, we initiated the discussion on groups generated by
involutions.  One subclass of such groups are called Coxeter groups,
whose definition includes one more condition: the entire group can be
specified once one knows the order of binary products of the
generators.  The generic presentation of Coxeter groups is therefore
of the form
\begin{eqnarray}
\left< r_i | (r_ir_j)^{q_{ij}} \right> \,, 
\label{pres}
\end{eqnarray}
where $i$ and $j$ run from 1 up to the number of generators.
Because the order of each generator is 2, we have the further
constraint that
\begin{eqnarray}
q_{ii} = 1 \qquad \forall i.
\end{eqnarray}
(Once again recall that we are nowhere using implied summation on
repeated indices.)  It can also be proved easily that
\begin{eqnarray}
q_{ij} = q_{ji} \,.
\end{eqnarray}
In order to specify a particular Coxeter group, one therefore needs
the following pieces of information:
\begin{enumerate}\itemsep=0pt
\item The number of generators.
\item The orders of binary products of the form $r_ir_j$ with $i<j$.
\end{enumerate}
If no power of the product of a particular pair of generators is equal
to the identity element, the corresponding $q_{ij}$ is taken to be
infinity. 

\begin{figure}
\begin{center}
\begin{tabular}{ll}
$A_N$ & \begin{picture}(100,10)
\lbullet 0,0,15,0/
\lbullet 15,0,30,0/
\lbullet 30,0,40,0/
\Text(50,0)[c]{$\cdots$}
\lbullet 70,0,60,0/
\lbullet 85,0,70,0/
\lbullet 100,0,85,0/
  \end{picture}
\\[2mm]
$B_N$ & \begin{picture}(100,10)
\lbullet 0,0,15,0/
\lbullet 15,0,30,0/
\lbullet 30,0,40,0/
\Text(50,0)[c]{$\cdots$}
\lbullet 70,0,60,0/
\lbullet 85,0,70,0/
\lbullet 100,0,85,0/
\Text(7.5,-3)[t]{\footnotesize 4}
  \end{picture}
\\[2mm]
$D_N$ & \begin{picture}(100,10)
\lbullet 0,0,15,0/
\lbullet 15,0,30,0/
\lbullet 30,0,40,0/
\Text(50,0)[c]{$\cdots$}
\lbullet 70,0,60,0/
\lbullet 85,0,70,0/
\lbullet 100,10,85,0/
\lbullet 100,-10,85,0/
  \end{picture}
\\[2mm]
$I_2(p)$ & \begin{picture}(20,10)
\lbullet 0,0,15,0/
\lbullet 15,0,0,0/
\Text(7.5,-3)[t]{\footnotesize $p$}
  \end{picture}
\\[2mm]
$F_4$ & \begin{picture}(50,10)
\lbullet 0,0,15,0/
\lbullet 15,0,30,0/
\lbullet 30,0,45,0/
\lbullet 45,0,30,0/
\Text(22.5,-3)[t]{\footnotesize 4}
  \end{picture}
\\[2mm]
$H_3$ & \begin{picture}(50,10)
\lbullet 0,0,15,0/
\lbullet 15,0,30,0/
\lbullet 30,0,15,0/
\Text(7.5,-3)[t]{\footnotesize 5}
  \end{picture}
\\[2mm]
$H_4$ & \begin{picture}(50,10)
\lbullet 0,0,15,0/
\lbullet 15,0,30,0/
\lbullet 30,0,45,0/
\lbullet 45,0,30,0/
\Text(7.5,-3)[t]{\footnotesize 5}
  \end{picture}

\end{tabular}
\end{center}

\caption{ Classification of irreducible finite Coxeter groups through
  Coxeter diagrams.  More Coxeter groups can be obtained by taking
  products of the groups given here.  We have not included the
  diagrams of $E_6$, $E_7$ and $E_8$, all of which have more than four
  generators and are therefore irrelevant for us.  Our nomenclature of
  the groups follows \textcite{davisbook}.  In each case, the
  subscript denotes the number of generators in the presentation of
  the group.  Products of the groups shown here qualify as Coxeter
  groups as well.}\label{f:finite}
\end{figure}
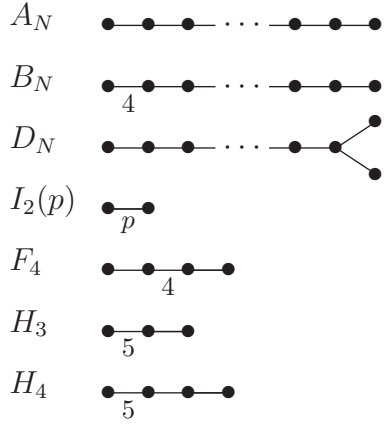
Coxeter diagrams constitute a pictorial way for depicting Coxeter
groups.  In a Coxeter diagram, each generator is depicted by a blob.
If $q_{ij}=3$ for some particular value of $i$ and $j$, then the
$i^{\rm th}$ and $j^{\rm th}$ blobs are joined by a line.  If
$q_{ij}>3$, there is still a line, but the value of $q_{ij}$ is
written above or below the line.  If $q_{ij}=2$, there is no line
joining the dots.  For example, the diagram
\begin{picture}(35,10)
\multiput(0,0)(15,0){3}{$\bullet$}
\put(3,3){\line(1,0){30}}
\end{picture}
would imply that the presentation of the Coxeter group is
\begin{eqnarray}
\left< r_1,r_2,r_3 \Big| r_1^2, r_2^2, r_3^2, (r_1r_2)^3, (r_2r_3)^3,
(r_1r_3)^2 \right> \,,
\label{prexamp}
\end{eqnarray}
where on the left we give a list of the generators and on the right we
give the combinations which are equal to the identity element.

 It has to be commented that the presentation of a group is
not unique.  For example, one can start with the elements $r_1$, $r_2$
and $r_3'=r_2r_3$ to generate the whole group given in \Eqn{prexamp}.
In this case, the generator $r_3'$ would be an element of order 3, and
the presentation of the group will be
\begin{eqnarray}
\left< r_1,r_2,r_3' \Big| r_1^2, r_2^2, (r_2r_3')^2, (r_1r_2)^3, r_3'^3,
(r_1r_2r_3')^2 \right> \,.
\end{eqnarray}
It will be hard to guess from this presentation that it is a Coxeter
group.  However, it must be, since the group is the same as that in
\Eqn{prexamp}.  Even the number of generators for the same group might
be different in two different presentations.  In our discussion, in
order to avoid such confusions, we will always talk about Coxeter
groups with the involution generators, and use the minimum number of
such generators necessary to write the presentation.  Thus, when we
talk about a group with two generators, we mean a group with two
involution generators unless something to the contrary is explicitly
stated.

In the present context, we will consider only Coxeter groups with 2 or
3 or 4 generators because, according to the restrictions put forth in
\Eqn{phasereln}, we can have at most two generators of the $S$ type
and two of the $T$ type.  Apart from a few exceptional groups, all 
finite Coxeter groups with connected Coxeter diagrams fall into one or
another of four infinite series, called $A_N$, $B_N$, $D_N$ and
$I_2(p)$ where the subscripts denote the number of generators.  In
\fig{f:finite}, we have shown the Coxeter diagrams for all four
series, and all exceptional groups whose number of generators is less
than or equal to 4.  Although the notations for the diagrams is more
or less universal, the names of the groups are not.  There are several
conventions, and we have followed the one of \textcite{davisbook}.

As a passing comment, note that the Coxeter groups $A_N$ are really
permutation groups.  Any permutation involving $n$ objects can be
generated by combining permutations which are transpositions of two
adjacent elements.  Thus, the group $S_n$ can be generated by
transpositions like $\tau_{1,2}$, $\tau_{2,3}$ and so on, up to
$\tau_{n-1,n}$.  It is also easy to see that products of any two
transposition will have order 2 if there are no common objects, and
order 3 if there is one common object.  Comparing with the diagram of
the $A_N$ groups, we therefore see that
\begin{eqnarray}
S_n = A_{n-1} \,.
\label{Sn}
\end{eqnarray}

From \fig{f:finite}, we see that the only irreducible finite Coxeter
groups whose number of generators is between 2 and 4 are the
following:
\begin{subequations}
\label{list}
\begin{eqnarray}
&& A_2, A_3, A_4; \\ && B_2, B_3, B_4; \\ && D_2, D_3, D_4; 
\\ && I_2(p) \qquad \mbox{with $p=3,4,\cdots$}; 
\\ && F_4, H_3, H_4 \,.
\end{eqnarray}
\end{subequations}
Some groups appear more than once in this list.  For example, a look
at \fig{f:finite} suggests the following equivalences:
\begin{subequations}
\label{equals}
\begin{eqnarray}
& I_2(3) = D_2 = A_2 \,,  \\ 
& D_3 = A_3 \,, \\
& I_2(4) = B_2 \,.
\end{eqnarray}
\end{subequations}
We will keep these coincidences in mind while identifying distinct
groups and discuss whether they give experimentally acceptable result
for the elements of the PMNS matrix.  In doing so, a few general
points should be kept in mind, which we summarize here.

Given any Coxeter diagram, we have to discuss the possibility of
assigning each of its dots to a generator of the type $S'_\alpha$ or
of the form $T_\ell$.  For this, we need to remember an important
result that can be proved easily: if $q_{ij}=2$ in \Eqn{pres} for some
specific values of $i$ and $j$, then $r_i$ and $r_j$ commute.  Now,
note that the $S'_\alpha$'s commute with one another.  Thus, if we
assign the generator $S'_1$ to one of the dots of a Coxeter diagram,
any other dot connected to it cannot be assigned to another $S'$
generator: it will have to be a $T$ generator.  For the same reason,
two connected dots cannot be assigned to two $T$ generators.

\subsection{Groups with two generators}\label{s:fc2}
From the discussion at the end of \sec{s:fcd}, it is clear that if we
look for a group with two generators, one of these generators must be
$S'_\alpha$ for some value of $\alpha$, and the other would be
$T_\ell$ for some $\ell$.  Therefore, we will be able to determine
only one element of the PMNS matrix, viz., the element
$U_{\ell\alpha}$.

The number of elements predicted might be disappointingly low, but
there is no reason to be unhappy about it, because we now argue that
none of these groups is viable so long as we use only the involution
generators.  Looking at \Eqn{equals}, we see that we need to consider
only $I_2(p)$ groups, because all other groups with two generators is
isomorphic to either $I_2(3)$ or $I_2(4)$.  However, the $I_2(p)$
groups are dihedral groups, i.e., symmetry groups of regular polygons,
as is obvious from the presentation implied by the Coxeter diagrams:
$\left< a,b | a^2, b^2, (ab)^p \right>$.  It is known that the dihedral
groups have only 1 or 2 dimensional irreducible representations.
Thus, they are unacceptable for us.

\subsection{Groups with three generators}\label{s:fc3}
With three generators, \fig{f:finite} shows us that we have the groups
$A_3$, $B_3$ and $H_3$.  The diagrams of these three groups look the
same: three dots joined by two links.  They differ only in the order
of one link.  Recalling that we cannot assign two $S'$ generators or
two $T$ generators to the ends of any link, we must alternate the two
types of generators.  Thus, we will either get the generators to be of
$S'TS'$ type, with whatever indices, or of the type $TS'T$.  In the
first case, we will be able to determine the absolute values of two
elements belonging to the same row of the PMNS matrix, whereas in
the second case we will be able to do the same for two elements
belonging to the same column.  In either case, we will be able to tell
the remaining element of the same column or row, as the case may be,
by using a suitable unitarity relation.   According to the
discussion of \sec{s:exp}, this extra value thus obtained must also be
of the form given in \Eqn{Usoln}, i.e., should be a solution of an
equation of the form \Eqn{sumcosS} or \Eqn{sumcosT}.

Because we obtain a full row or a full column, it is not necessary to
check whether the group has any 3-dimensional irrep.  If the only
irreps are 1 and 2-dimensional, each row and each column must have at
least one element equal to zero.  Thus, if we just determine one row
or one column and do not find a zero in it, it guarantees that the
representation is not reducible.  On the other hand, if we obtain at
least one zero, it does not guarantee that the representation is
reducible, but it does say that the group should be discarded because
none of the entries in \Eqn{Usqlim} is consistent with zero.

\begin{table}
\caption{Allowed generators for the group $A_3$.   (Note: This
  is the Coxeter nomenclature.  This group is more commonly known as
  $S_4$, and the GAP identifier of this group is \gap 24
  12/.)}\label{t:A3}

$$
\begin{array}{l|c|c|c}
\multirow{2}{*}{\mbox{Generators}} & \multicolumn2{c|}{\mbox{Determination
    from}} & \mbox{Which solution} \\ \cline{2-3}
& \mbox{\Eqn{Usoln}} &  \mbox{Unitarity} & \mbox{of \Eqn{2solns}} \\ 
\hline
\displaystyle{\{S'_1, T_\mu,  S'_2\} \atop \{S'_2, T_\mu,
  S'_1\} }
& |U_{\mu1}|^2 = |U_{\mu2}|^2 = \frac14 & |U_{\mu3}|^2 = 
\frac12 & N=12 \\[5mm]
\displaystyle{ \{S'_1, T_\tau,  S'_2\} \atop \{S'_2, T_\tau,
  S'_1\} } 
& |U_{\tau1}|^2 = |U_{\tau2}|^2 = \frac14 & |U_{\tau3}|^2 = 
\frac12 & N=12 \\[5mm]
\displaystyle { \{T_e,  S'_2, T_\mu\}  \atop \{T_\mu,  S'_2,
  T_e\} } 
& |U_{e2}|^2 = |U_{\mu2}|^2 = \frac14 &
|U_{\tau2}|^2 =  
\frac12 & N=12 \\[5mm] 
\displaystyle { \{T_e,  S'_2, T_\tau\} \atop \{T_\tau,  S'_2,
  T_e\} }
& |U_{e2}|^2 = |U_{\tau2}|^2 = \frac14 &
|U_{\mu2}|^2 =  
\frac12 & N=12 \\
\end{array}
$$
\end{table}
\paragraph{The group $A_3$:}  In this case, both links have order 3.
Looking at \Eqn{allowed}, we see that the choice of $k$ is unique.
For each possible choice of the generators, we give the list of matrix
element whose absolute values are determined through \Eqn{aUcox}, and
both these values should be $\frac14$ as shown in \Eqn{allowed}.
Unitarity of the mixing matrix would dictate that the absolute square
of the remaining element of the same row or same column should be
equal to $\frac12$, as discussed above.  We look for all combinations
of generators and list the acceptable ones in \tabl{t:A3}.

This group is isomorphic to the permutation group $S_4$, as pointed
out in \Eqn{Sn}, and has been encountered by various authors earlier.
However, this does not necessarily mean that their analysis is same as
that of ours, or that they obtain the same values of the PMNS elements
as we do.  As we commented before, the presentation of any group is
not unique.  Some authors \cite{Brown:1984dk, Mohapatra:2003tw,
  Hagedorn:2006ug}, while considering the group $S_4$ in models
inspired by grand unified theories, did not take all the
generators of $S_4$ to be involutions and obtained different results.
In contrast, we consider only involution generators.  There are also
computer searches \cite{Lavoura:2014kwa} of finite groups with three
generators.  We will discuss their results shortly.

\begin{table}
\caption{Allowed generators for the group $B_3$.   (Note: This is the
  Coxeter nomenclature.  The group has GAP identifier \gap 48 48/, and
  is isomorphic to $A_3 \times Z_2$ or $S_4 \times Z_2$
  \cite{Parattu:2010cy}.)}\label{t:B3}

$$
\begin{array}{l|c|c|c}
\multirow{2}{*}{\mbox{Generators}} & \multicolumn2{c|}{\mbox{Determination
    from}} & \mbox{Which solution} \\ \cline{2-3}
& \mbox{\Eqn{Usoln}} &  \mbox{Unitarity} & \mbox{of \Eqn{2solns}} \\  
\hline
\{T_\mu,  S'_2, T_\tau\} & 
|U_{\mu2}|^2 = \frac12, |U_{\tau2}|^2 = \frac14 & |U_{e2}|^2 =  
\frac14  & N=12 \\[2mm]
  \{T_\tau, S'_2, T_\mu\} &
|U_{\tau2}|^2 = \frac12,  |U_{\mu2}|^2 = \frac14 & |U_{e2}|^2 =  
\frac14 & N=12 \\[2mm]
\{S'_3, T_\mu,  S'_1\} 
& |U_{\mu3}|^2 = \frac12, |U_{\mu1}|^2 = \frac14 & |U_{\mu2}|^2 = 
\frac14 & N=12 \\[2mm] 
\{S'_3, T_\mu,  S'_2\} 
& |U_{\mu3}|^2 = \frac12, |U_{\mu2}|^2 = \frac14 & |U_{\mu1}|^2 = 
\frac14 & N=12 \\[2mm] 
\{S'_3, T_\tau,  S'_1\} 
& |U_{\tau3}|^2 = \frac12, |U_{\tau1}|^2 = \frac14 & |U_{\tau2}|^2 = 
\frac14 & N=12 \\[2mm] 
\{S'_3, T_\tau,  S'_2\} 
& |U_{\tau3}|^2 = \frac12, |U_{\tau2}|^2 = \frac14 & |U_{\tau1}|^2 = 
\frac14 & N=12 \\[2mm]
\{T_\mu,  S'_2, T_e\} & 
|U_{\mu2}|^2 = \frac12, |U_{e2}|^2 = \frac14 & |U_{\tau2}|^2 =  
\frac14  & N=12 \\[2mm]
\{T_\tau,  S'_2, T_e\} & 
|U_{\tau2}|^2 = \frac12, |U_{e2}|^2 = \frac14 & |U_{\mu2}|^2 =  
\frac14  & N=12 \\[2mm]
\end{array}
$$
\end{table}
\paragraph{The group $B_3$:}  Here, one link has order 4 and one has
order 3.  As in \fig{f:finite}, we take the left link to have order 4.
Because of this link of order 4, this time we will have to distinguish
which generator corresponds to the right dot and which one to the
left, something that was not important for the previous case.  We list
all allowed possibilities in \tabl{t:B3}.  This group appeared in the
discussion of \textcite{Grimus:2006wy}, but only as a factor in the
semi-direct product $(Z_2 \times Z_2 \times Z_2) \rtimes B_3$.

\paragraph{The group $H_3$:}  As in \fig{f:finite}, we take the left
link to have order 5.  The right link then has order 3.  The modulus
squared value of the matrix element corresponding to the right link is
$\frac14$, according to \Eqn{allowed}.  The value coming from the left
link can be either $\frac18(3+\surd5)=0.6545$ or
$\frac18(3-\surd5)=0.0955$, depending on the value of
$k_{\ell\alpha}$.  In either case, the modulus squared values of the
three elements in one row or one column should be $\frac18(3+\surd5)$,
$\frac18(3-\surd5)$ and $\frac14$, in whatever order.  There is only
one element in \Eqn{Usqlim} that can be as big as $0.6545$, and that
is $|U_{e1}|^2$.  In the first row, there is no choice of generators
which can give rise to those values in \tabl{t:H3}.  The first column,
however, is acceptable, and we show the choice of generators which can
give rise to those values. This group is isomorphic to direct product
of alternating group ${\cal A}_5$ (not to be confused with the Coxeter
group $A_5$) and $Z_2$.
\begin{table}
\caption{Allowed generators for the group $H_3$.   (Note: This group
  has GAP identifier \gap 120 35/.)}\label{t:H3}

$$
\begin{array}{l|c|c|c}
\multirow{2}{*}{\mbox{Generators}} & \multicolumn2{c|}{\mbox{Determination
    from}} & \mbox{Which solution} \\ \cline{2-3}
& \mbox{\Eqn{Usoln}} &  \mbox{Unitarity} & \mbox{of \Eqn{2solns}} \\  
\hline
\{T_e,  S'_1, T_\mu\} & 
|U_{e1}|^2 = 0.6545, |U_{\mu1}|^2 = \frac14 & |U_{\tau1}|^2 =  
0.0955 & N=15 \\[2mm]
\{T_e,  S'_1, T_\tau\} & 
|U_{e1}|^2 = 0.6545, |U_{\tau1}|^2 = \frac14 & |U_{\mu1}|^2 =  
0.0955 & N=15 \\[2mm]
\{T_\mu, S'_1, T_\tau\} & 
|U_{\mu1}|^2 = 0.0955, |U_{\tau1}|^2 = \frac14 & |U_{e1}|^2 =  
0.6545 & N=15 \\[2mm]
\{T_\tau,  S'_1, T_\mu\} & 
|U_{\tau1}|^2 = 0.0955, |U_{\mu1}|^2 = \frac14 & |U_{e1}|^2 =  
0.6545 & N=15 
\end{array}
$$
\end{table}

\paragraph{Discussion of previous work:} Fixing a column or a row of
the PMNS matrix by considering three involution generators was
considered  by Lavoura and
Ludl\cite{Lavoura:2014kwa}.  They made computer searches in the GAP
\cite{GAP4} database.  This database lists finite groups, including
their properties and representations.  Each group is denoted by two
numbers in the form $[a,b]$.  The first number is the cardinality of
the group, and the second one is an arbitrary serial number assigned
to groups of equal cardinality.  These GAP identifiers have been
mentioned in the captions of the tables for the groups discussed
above.

For $N=12$ solution of \Eqn{2solns}, the smallest group found by
Lavoura and Ludl is \gap 24 12/ which is nothing but the $A_3$ group.
We also got the same solution for the $B_3$ group whose cardinality is
48. This is a bigger group containing the $A_3$ group.  By the
definition of the group $B_3$,
\begin{eqnarray}
(T_\mu S'_2)^4 = (T_\tau S'_2)^3 =1 \,,
\label{B3orders}
\end{eqnarray}
which helps us determine two elements of the second column of the PMNS
matrix.  For the remaining element of the same column, we can deduce 
\begin{eqnarray}
(T_e S'_2)^3 &=& (S'_2 T_e)^3,
\end{eqnarray}
by using relations like those in \Eqn{Tprod} and 
\begin{eqnarray}
T_\tau S'_2 T_\tau &=&  S'_2 T_\tau  S'_2 \,, \nonumber\\* 
T_\mu S'_2 T_\mu &=& S'_2 T_\mu  S'_2 T_\mu  S'_2 \,, 
\end{eqnarray}
which follow from \Eqn{B3orders}.  From this, it is trivial to show
that the sixth power of $T_e S'_2$ or of $S'_2 T_e$ is equal to the
identity element.  Thus, when one chooses the lowest value, i.e., 3,
for $p_{e2}$, the order of $T_eS'_2$, then because of an extra
relation, instead of the $B_3$ group a subgroup of that group is
really being considered.  And that subgroup is $A_3$.  Thus $B_3$ has
twice as many elements as $A_3$.  In the GAP \cite{GAP4} database,
this group is called \gap 48 48/.

For the $N=15$ case, the smallest group they found has SmallGroup Id
as $\gap 60 5/$ whereas we found $H_3$, whose cardinality is 120 and
GAP \cite{GAP4} database Id \gap 120 35/, as the smallest group with
the $N=15$ solution.  This is because we consider only Coxeter groups,
and $\gap 60 5/$, a subgroup of $H_3$, is not a Coxeter group.
Lavoura and Ludl \cite{Lavoura:2014kwa} found some bigger groups as
well, but we suspect that those solutions involve non-involution
generators.

\subsection{Groups with four generators}\label{s:fc4}
From \fig{f:finite}, we see that the finite groups with four
generators are $A_4$, $B_4$, $D_4$, $F_4$ and $H_4$.  Let us give the
final result first: none of these groups is allowed.  We explain the
reasons in what follows.

Let us discuss $D_4$ first.  Its Coxeter diagram has one nodal blob,
from which three links come out to meet the three other blobs.  If we
assign an $S'$ generator to the nodal blob, we must assign $T$-type
generators to all other blobs.  However, there are only two $T$-type
independent generators, so this group is untenable.  Even if we
venture to put three different $T$-type generators on the three blobs,
unitarity condition is not fulfilled since we get $|U_{\ell\alpha}|^2
= \frac14$ for all elements in a column.  The argument with a $T$-type
generator as the middle blob is the same, and need not be repeated.

For all other groups the list, the Coxeter diagrams are linear.
Because the $S'$-type and $T$-type generators alternate on a line, one
of the two extreme blobs must correspond to an $S'$-type generator and
the other one to a $T$-type generator.  There is no line joining these
two blobs, which means that there is a $W_{\ell\alpha}$ which is of
order 2.  But $p_{\ell\alpha}=2$ does not give any acceptable
solution, as noted in \Eqn{Usoln}.

Let us now ignore the constraint that all the calculated PMNS elements
must comply with the experimentally observed values within a $3\sigma$
limit and examine whether any of these groups can even satisfy the
unitarity conditions.  There is at least one pair of $T$ and $S'$
generators which are not connected by a link, i.e., the corresponding
$W_{\ell\alpha}$ has $p_{\ell\alpha}=2$.  Therefore, one entry of the
calculated PMNS matrix must be zero or unity, depending on the value
of $k_{\ell\alpha}$.  If $k_{\ell\alpha}=0$ and therefore
$|U_{\ell\alpha}|^2=1$ for one element, the other two in the same row
and in the same column must be zero, and so there will be one
eigenstate that will not mix with the other two.  On the contrary, if
$k_{\ell\alpha}=1$, that particular element vanishes, and so the sum
of absolute squared values of two other elements of the PMNS matrix in
the same row  must be unity. This is true also for the column containing this 
zero entry.

For the $A_4$ group, $p_{\ell\alpha}=3$ for all the links and
$|U_{\ell\alpha}|^2=\frac14$.  Thus two such entries cannot add up to
1, and one obtains a contradiction since the matrix should be unitary.
The group $H_4$ has $p_{\ell\alpha}=3,5$.  There are no two allowed
values of $|U_{\ell\alpha}|^2$, corresponding to these values of
$p_{\ell\alpha}$, which add up to 1, and hence the fate of this group
is the same.  For the groups $B_4$ and $F_4$, one link has values of
$p_{\ell\alpha}=4$, which can give $|U_{\ell\alpha}|^2=\frac12$, but
then the other two have $p_{\ell\alpha}=3$, which can produce a
maximum of $\frac14$.  Thus we'll get at most $\frac34$ as the
required sum.  So we see that none of the 4-generator groups can
satisfy the constraints of unitarity: we meet a contradiction.

\subsection{Direct product groups}\label{s:fcp}
So far, the groups we have discussed may be called {\em irreducible
  Coxeter groups}.  We can also consider reducible ones, i.e., groups
which are direct products of more than one Coxeter groups.  If first
we look at groups with four generators, we have the following options:
\begin{eqnarray}
\begin{array}{r@{\qquad}l}
\mbox{(a)} & A_1 \times X_3 \,,  \\ 
\mbox{(b)} & \Big( A_1 \Big)^2 \times I_2(p) \,, \\
\mbox{(c)} & \Big( A_1 \Big)^4 \,, \\ 
\mbox{(d)} & I_2(p)\times I_2(p') \,,
\end{array}
\label{product}
\end{eqnarray}
where $X_3$ can be any Coxeter group with three connected generators. 
For any of the choices of $X_3$, the Coxeter diagram will consist of
one isolated blob, and the other three arranged in a manner shown in
\fig{f:finite}.  Suppose the isolated one represents a $T$-type
generator.  Then it is not connected with any of the $S'$ generators,
implying that two elements in a row are zero.  If the isolated one is
an $S'$ generator, then two elements in a column are zero.  None of
these options is acceptable.  The next two options will also have at
least one generator which is not connected to any other generator, and
therefore gives zeros in the PMNS matrix.

If we take the last option of \Eqn{product}, then the Coxeter diagram
will have two disconnected parts, each with two blobs.  Each
disconnected part will have to contain one $T$-type and one $S'$-type
generator.  Here again, each $T$ blob will not be linked with one $S'$
blob that is in the other connected part, and therefore there should
be two zeros in the PMNS matrix.

Groups with two or three generators will have the same problem with
disconnected parts in the Coxeter diagram, and are untenable.

\section{More involution groups}\label{s:inv}
We see that, among the finite Coxeter groups, the only acceptable
solutions have three generators, which give only one column 
or one row of the PMNS matrix.  Since we could not find any group with
four generators, we cannot determine the entire PMNS matrix.   Because of
this, we now look for other options.  We recall that our analysis is
based on the relations derived in \sec{s:deg} and \sec{s:exp},
especially on the formula for PMNS matrix elements in \Eqn{Usoln} and
its correspondence with experimental data in \Eqn{allowed}.  It should
be noticed that none of these key formulas depend on the Coxeter
nature of the group.  In fact, the definition of Coxeter groups was
not even given until \sec{s:fc}.  Rather, these formulas depend on the
fact that all generators of the group are involutions.  Therefore, we
take this opportunity to explore other groups which are generated by
involutions, but whose presentation is not of the form given in
\Eqn{pres}, and which therefore do not qualify as Coxeter groups.

To this end, we employ the following strategy.  We start with
\Eqn{n/N} which also depends only on the fact that the group is
generated by involutions.  Irrespective of the experimental data, we
first analyze how many solutions to this equation can be obtained, and
we outline some criteria for selecting a subset of them.  Then we make
computer searches for these solutions, using four involution
generators. 
 
First, let us look at the solutions of \Eqn{n/N}.  There can be
various solutions which involve vanishing elements of the PMNS matrix.
We disregard them and enumerate only the solutions where none of the
PMNS elements is zero.  
\begin{enumerate}

\item The solution with $N=12$ presented in \Eqn{2solns} can be used
  for each row and each column, producing the mod-squared values of
  the elements of a unitary matrix:
\begin{subequations}
\label{cases}
\begin{eqnarray}
|U^2|_{12} = \left(
\begin{array}{ccc}
\frac12 & \frac14 & \frac14\\
\frac14 & \frac12 & \frac14 \\
\frac14 & \frac14 & \frac12\\
\end{array}
\right) 
\label{12only}
\end{eqnarray}
Of course it will have to be understood that variations can be
obtained by reshuffling  rows or columns.

\item One can also use the solution with $N=15$ from \Eqn{2solns} to
  construct the following values of the mod-squared values of the
  elements:
\begin{eqnarray}
|U^2|_{15} = \left(
\begin{array}{ccc}
\frac18(3+\surd5) & \frac14 & \frac18(3-\surd5)\\
\frac18(3-\surd5) & \frac18(3+\surd5) & \frac14 \\
\frac14 & \frac18(3-\surd5) & \frac18(3+\surd5)\\
\end{array}
\right) 
\label{15only}
\end{eqnarray}
As before, reshuffling of rows or columns is allowed.

\item There can also be solutions where some rows will come from the
  $N=12$ solution and some from the $N=15$ solution.  Of course, the
  latter kind must come in pairs because the occurrence of the
  irrational value of either $\cos^2(\pi/5)$ or $\cos^2(2\pi/5)$
  cannot be consistent with the unitarity condition without the
  occurrence of the other.  Thus, the solution is
\begin{eqnarray}
|U^2|_{12\&15} = \left(
\begin{array}{ccc}
\frac18(3+\surd5) & \frac14 & \frac18(3-\surd5)\\
\frac18(3-\surd5) & \frac14 & \frac18(3+\surd5)\\
\frac14 & \frac12 & \frac14 \\
\end{array}
\right) \,.
\label{1215}
\end{eqnarray}
\end{subequations}
Here also, reshuffling of rows or columns produces acceptable
alternatives.
\end{enumerate}

Note that none of the matrices shown in \Eqn{cases} is consistent with
the experimental data for all elements of the PMNS matrix.  The 
agreement is  very bad  for the form shown in
\Eqn{15only}, where no row or no column is fully consistent with the
data.  The form of \Eqn{12only} can be consistent with the data for
the lower two rows of the PMNS matrix.  In the top row, it misses one
element by a small amount, and does not agree for the other two
elements within the $3\sigma$ limits.  The solution of \Eqn{1215} is
definitely consistent with the values of five elements, and among the
remaining ones, it misses one very narrowly.  With little to choose
from between these two forms, we perform the search for all three
forms, showing that even these forms are very difficult to obtain.

The search is made by using GAP \cite{GAP4}.  We take four involution
generators, since that will determine the  absolute
values of all elements of the  PMNS matrix.  For the sake of
definiteness, let us call these four generators $T_e$, $T_\mu$,
$S'_1$, $S'_2$.  We feed a presentation involving these generators of
the following form and obtain the group:
\begin{eqnarray}
\bigg< T_e, T_\mu, S'_1, S'_2 & \Big| & T_e^2, T_\mu^2, S_1'^2, S_2'^2,
(T_e T_\mu)^2, (S_1' S_2')^2, \nonumber\\* 
&& (T_e S_1')^{P_{e1}}, (T_e S_2')^{P_{e2}}, (T_e S_3')^{P_{e3}}, \nonumber\\* 
&& (T_\mu S_1')^{P_{\mu1}}, (T_\mu S_2')^{P_{\mu2}}, (T_\mu
S_3')^{P_{\mu3}}, \nonumber\\* 
&& (T_\tau S_1')^{P_{\tau1}}, (T_\tau S_2')^{P_{\tau2}}, (T_\tau
S_3')^{P_{\tau3}} \bigg> \,,
\label{4gengp}
\end{eqnarray}
where $T_\tau = T_e T_\mu$ and $S_3' = S_1' S_2'$.  It should be noted
that $T_\tau$ and $S'_3$ are {\em not} generators of the group: these
notations have been introduced only for the sake of brevity.

The first four conditions of \Eqn{4gengp} indicate that all generators
are involutions.  The next two conditions state that $T_e$ commutes
with $T_\mu$ whereas $S_1'$ commutes with $S_2'$.  These six
conditions of the presentation should be the same for all searches.

Each of the other conditions announces the order of one group element.
If we had had only the orders of the elements $T_e S_1'$, $T_e S_2'$,
$T_\mu S_1'$ and $T_\mu S_2'$, and if these orders had corresponded to
the orders obtainable from the Coxeter diagrams of \fig{f:finite} with
four generators, the resulting group would have been a finite Coxeter
group.  However, we exclude that possibility by making some
exceptions, as we explain now.

Suppose, for the sake of definiteness, that we are trying to see how
one might obtain the PMNS matrix of the form shown in \Eqn{12only}.
In each row, one of the elements is equal to $\frac12$, which must
come from $k/p=\frac14$.  Previously, we used \Eqn{gcd} to conclude
that therefore these elements should have order $p=4$.  But now we
allow for multiples of 4.  As long as we keep the same value of $k/p$,
we can obtain the same value of $|U_{\ell\alpha}|^2$.  For this
reason, we write the order of the element with a uppercase $P$ in
\Eqn{4gengp}, indicating that $k$ and $p$ need not be relatively
prime.  In our search, we allow for a GCD up to 3 for each pair of $k$
and $p$.

Even after making that adjustment, if we restrict the presentation
table of \Eqn{4gengp} only to the first six declarations and the
orders of $T_e S_1'$, $T_e S_2'$, $T_\mu S_1'$ and $T_\mu S_2'$ only,
we would obtain a Coxeter group because of the definition given
in \Eqn{pres}.  The group would be infinite, because it would not
correspond to any of the diagrams of \fig{f:finite}.  But we have more
conditions.  We take these other conditions in such a way that they
provide us with an acceptable value of $|U_{\ell\alpha}|^2$ that can
satisfy \Eqn{n/N}.

Let us explain the strategy in some more detail.   Suppose we want to
obtain the PMNS matrix of the form given in \Eqn{12only}, with the top
left element equal to $\frac12$.  As said before, it needs
$k/p=\frac14$.  Since $k$ is an integer, it means that $p$ must be a
multiple of 4, so we denote it by $4a_1$, where $a_1$ is an integer.
Similarly, since the other elements of this row should come from
$k/p=\frac13$, we can denote them by $3a_2$ and $3a_3$, where $a_2$
and $a_3$ are integers.  Proceeding in this fashion, we obtain the
following forms for the orders $P_{\ell\alpha}$ of the group 
elements responsible for different elements of the PMNS matrix: 
\begin{subequations}
\begin{eqnarray}
\left[ \begin{array}{ccc} 
P_{e1} & P_{e2} & P_{e3} \\
P_{\mu1} & P_{\mu2} & P_{\mu3} \\
P_{\tau1} & P_{\tau2} & P_{\tau3} \\
  \end{array} \right]_{\tiny\rm\Eqn{12only}}
= \left[ \begin{array}{ccc}
4a_1 & 3a_2 & 3a_3 \\
3b_1 & 4b_2 & 3b_3 \\
3c_1 & 3c_2 & 4c_3 \\
  \end{array} \right] \,,
\\ 
\left[ \begin{array}{ccc} 
P_{e1} & P_{e2} & P_{e3} \\
P_{\mu1} & P_{\mu2} & P_{\mu3} \\
P_{\tau1} & P_{\tau2} & P_{\tau3} \\
  \end{array} \right]_{\tiny\rm\Eqn{15only}}
= \left[ \begin{array}{ccc}
5a_1 & 3a_2 & 5a_3 \\
5b_1 & 5b_2 & 3b_3 \\
3c_1 & 5c_2 & 5c_3 \\
  \end{array} \right] \,,
\\
\left[ \begin{array}{ccc} 
P_{e1} & P_{e2} & P_{e3} \\
P_{\mu1} & P_{\mu2} & P_{\mu3} \\
P_{\tau1} & P_{\tau2} & P_{\tau3} \\
  \end{array} \right]_{\tiny\rm\Eqn{1215}}
= \left[ \begin{array}{ccc}
5a_1 & 3a_2 & 5a_3 \\
5b_1 & 3b_2 & 5b_3 \\
3c_1 & 4c_2 & 3c_3 \\
  \end{array} \right] \,,
\end{eqnarray}
\end{subequations}
For each case, it has to be remembered that we allow for reshuffling
of rows and columns.  For a search in each category, we take the value
of each of the integers to run from 1 to 3.

\begin{table}
\caption{Result of the search described in the text.  For
  each kind of PMNS matrix elements, we give the possible finite
  subgroup of a Coxeter group.  The numbers given under the columns
  marked ``cardinality'' and ``serial Id'', taken together, form what
  has been called the {\tt SmallGroup} Id in the GAP internet
  archives \cite{GAP4}.}\label{t:subgrps}

\begin{center}
\begin{tabular}{ccccl}
\hline
Type of & \multicolumn{2}{c}{Finite subgroup} & \multirow{2}{*}{3-d
  irrep?}  & \multirow{2}{*}{Remarks}  
\\ \cline{2-3}
 PMNS matrix & Cardinality & Serial Id & \\ 
\hline
\multirow{3}{*}{\Eqn{12only}} & 12 & 4 & No & Hexagon dihedral group \\
& 108 & 17 & No \\
& 576 & 8654 & No \\ \hline 
\Eqn{15only} & \multicolumn{3}{c}{(no solution)} 
\\ \hline
\Eqn{1215} & 1080 & 260 & Yes & $\Sigma(360 \times 3)$ \\ 
\hline
\end{tabular}
\end{center}
\end{table}
Because there are extra conditions now, we do not obtain a Coxeter
group.  Rather, we obtain a subgroup of a Coxeter group that will be
obtained by deleting all conditions involving $T_\tau=T_e T_\mu$ and
$S'_3=S'_1 S'_2$ from \Eqn{4gengp}, which will be an infinite group
because it does not appear in \fig{f:finite}.  Using GAP \cite{GAP4},
we search whether the resulting group with any given choice of the
integers $a_1$ to $c_3$ is finite.  We limit our search to groups with
number of elements between 9 and 1200.  Smaller groups are not
considered because they do not have any 3-dimensional irrep, and
larger groups are probably less interesting because they are not very
economical.  Whenever we find a finite group with cardinality in this
range, we announce it in \tabl{t:subgrps}.  And finally, also from
GAP, we find whether the group obtained has any 3-dimensional
irreducible representation.  The result is also shown in
\tabl{t:subgrps}.

We see that, for groups with four involution generators and with
cardinality up to 1200, we obtain no solution of the form advocated in
\Eqn{15only}.  For the form given in \Eqn{1215}, we find one solution
in the same range.  This group has the GAP Id \gap 1080 260/, and is
the $\Sigma(360\times3)$ group \cite{Hagedorn:2013nra} as shown in
\tabl{t:subgrps}.  For the form given in \Eqn{12only}, we find some
groups, but none of them has any 3-dimensional irrep.  In fact, in
desperation we continued the search somewhat further for this case and
found that the next available finite group has 1728 elements, but this
one also does not have any 3-dimensional irrep.  We did not continue
further.

\section{Summary}\label{s:out}
We have considered all finite Coxeter groups whose number of
generators range from 2 to 4 which appear in \fig{f:finite}, 
as well as their direct products.  We have identified the involution
generators with residual symmetry generators in the low energy
leptonic sector, and tried to see if they give an acceptable PMNS
matrix.  It would have been great if we could find a group with 4
generators to satisfy all known experimental constraints, because then
we could have predicted the magnitudes of all elements of the PMNS
matrix.  Unfortunately, our analysis shows that such groups are not
compatible with experimental results, in concurrence with results
obtained by earlier authors who have considered many of the groups
discussed in the present work.  There are acceptable solutions with
Coxeter groups with 3 generators, although these are only partial
solutions.  With a 3 generator group, we can calculate the magnitudes
of elements in one row or in one column of the PMNS matrix.  Such
solutions have been shown in \sec{s:fc3}.  For a group with 2
generators, only one element can be calculated.  However, no such
group is acceptable, for reasons described in \sec{s:fc2}.

We then searched  for other groups that can be generated by 4 involution
generators.  We identify the very few types of PMNS matrices that can
be generated from any involutionary group, and tried to see whether
any involutionary group with 4 generators can provide that pattern.
Within the limits in which we have performed the search, we have
obtained a few such subgroups, which we have listed in
\tabl{t:subgrps}.  Some of these groups do not have any 3-dimensional
irreducible representation and are therefore unacceptable.  There is
only one solution that is consistent with non-zero entries for each
element, and has 3-dimensional irrep as well.

One of our working assumptions was that the determinant of each 
generator is equal to 1, as announced in \Eqn{det=1}.  For $Z_2$
generators, the only other possibility is to have determinant equal to
$-1$.  If we allow this possibility as well, the argument leading to
\Eqn{n/N} will be modified.  For products of the form $T_\ell
S'_\alpha$ which will have negative determinant, the cosines appearing
in \Eqn{n/N} will be replaced by sines.  Using $\sin x =
\cos(\frac12\pi-x)$, we can turn these equations into equations
involving cosines again, just like \Eqn{n/N}.  Thus we will not get
any new solutions.

\paragraph*{Acknowledgements:} We thank Uday Shankar Chakraborty for
pointing out at some references on Coxeter groups, to Max Horn for
suggestions regarding the use of the computer program GAP, and to
Renato Fonseca for pointing out some important works in the field
which we had missed in the first version of this paper submitted to
the archives (\url{xxx . lanl . gov}).

\paragraph*{Note added:} Even after the appearance of this work in the
archives \url{xxx . lanl . gov}, new results on GAP searches have
appeared, e.g., in \textcite{Jurciukonis:2016wrh}.

\small

\bibliographystyle{JHEP}
\bibliography{coxeter}

\end{document}